\documentclass[12pt]{article}

\usepackage{graphicx} 
\usepackage{hyperref}

\hypersetup{
    colorlinks = true,
   linktocpage = true,
    citecolor = {blue}
}

\usepackage{algpseudocode}
\usepackage{subcaption}
\usepackage{svg}

\usepackage{latexsym}
\usepackage{amsmath}
\usepackage{amssymb}
\usepackage{mathrsfs}
\usepackage{amsfonts}
\usepackage{eucal}
\usepackage{amsthm}
\usepackage{graphicx}
\usepackage{color}
\usepackage{youngtab}
\usepackage{slashed}
\usepackage{tikz-cd}
\usepackage{calc}

\usepackage{accents}

\newcommand {\qe}{\mathfrak{q}}

\newcommand {\CalA} {\mathcal A}

\newcommand {\CalC} {\mathcal C}

\newcommand {\CalD} {\mathcal D}

\newcommand {\CalN} {\mathcal N}
\newcommand {\CalO} {\mathcal O}

\newcommand {\CalX} {\mathcal X}

\newcommand {\CalZ} {\mathcal Z}

\newcommand {\BC}   {\mathbb C}

\newcommand {\BN}   {\mathbb N}

\newcommand {\BR}   {\mathbb R}
\newcommand {\BP}   {\mathbb P}

\newcommand {\BZ}   {\mathbb Z}

\newcommand {\ve}{\varepsilon}

\newcommand {\vt}{\vartheta}

\newcommand{\bla}{\boldsymbol{\lambda}}

\newcommand{\ii}{\mathrm{i}}

\newcommand{\beq}{\begin{equation}}
\newcommand{\eeq}{\end{equation}}

\title{Noncommutative Jacobi identity, \\
and gauge theory}

\author{Andrei Grekov, Nikita Nekrasov\\
Simons Center for Geometry and Physics$^{n}$\\
Yang Institute for Theoretical Physics$^{g,n}$\\ 
Stony Brook University, Stony Brook NY 11794-3636, USA}
\date{}

\begin{document}

\maketitle

\begin{abstract}
We prove the noncommutative analogue of Jacobi triple product identity. As an application we organizing the $q$-characters of ${\hat A}_{r}$-type quiver gauge theories into an infinite product. We conjecture the gauge origami theory interpretation of the Jacobi identity. 

\end{abstract}
\tableofcontents 

\section{Introduction}

Four dimensional gauge theories connect the topology of moduli spaces of instantons to classical and quantum integrable systems, theory of isomonodromic deformations, and two dimensional conformal field theory. This correspondence, named the  BPS/CFT correspondence \cite{BPSCFT}, is based on the observation \cite{Nekrasov:2002qd}
that the six dimensional 
$(2,0)$ theory, the theory of fivebranes in M-theory/IIA string, can be used as a sort of infinite-dimensional Hecke correspondence connecting four and two dimensional physics. The celebrated example of the BPS/CFT correspondence is the AGT conjecture \cite{AGT}. Not all aspects of BPS/CFT are simple in the six dimensional perspective. A powerful tool, the gauge origami \cite{BPSCFT3}, is motivated by the IIB string perspective. 

In this paper we explore one of the instances of the BPS/CFT correspondence, illustrating the fruitfulness of the 
multi-dimensional perspective. We shall prove the noncommutative analogue of Jacobi triple product formula. 
The latter can be understood both purely combinatorially, or through the lens of boson-fermion correspondence
in two dimensional (conformal) field theory. Our generalization of the formula might look quite unexpected to the two dimensional observer, however we shall argue it is quite natural in the four dimensional gauge theory framework. 

The paper is organized as follows. In the section $\bf 2$ we present and prove the noncommutative Jacobi identity and the accompanying bi-linear identity, somewhat similar to Hirota identity. 
In section $\bf 3$ we very briefly sketch some of the applications of the noncommutative Jacobi identity. 
In section $\bf 4$ we explain the gauge theory setup which might be useful in trying to case the identity in geometric framework. 

In the accompanying papers \cite{NGtoap, NGhighertimes} we use the identities presented in this paper to construct the Lax operator of elliptic Calogero-Moser system,  its eigenvectors, as well as the related isomonodromic flat connection and its horizontal sections. We also generalize this construction to all $A$-type quiver gauge theories
in four dimensions. We shall also derive the Seiberg-Witten \cite{SW} geometry of these theories ${\CalN}=2$ chirally deformed by the higher Casimirs operators in the ultraviolet. 

{\bf Acknowledgements}.  We thank A.~Okounkov and Y.~Soibelman for encouraging discussions. One of us (NN) thanks IHES for hospitality while part of this work has been done. Research is partly supported by NSF PHY Award 2310279.

\section{Noncommutative identities}

\subsection{Jacobi identity and bosonization}

Consider the product
\beq
{\CalZ}(z, {\qe}) = \prod_{r> 0} \left( 1 + z {\qe}^{r} \right) \left( 1 + z^{-1} {\qe}^{r} \right)
\eeq
over the set of positive half integers, $r \in {\BZ} + \frac 12$. It is well-known from quantum statistical mechanics that this product decomposes as:
\beq
{\CalZ}(z, {\qe}) = \sum_{M \in {\BZ}, {\lambda} \in {\Lambda}} z^{M} {\qe}^{\frac{M^2}{2} + |{\lambda}|}\, . 
\label{eq:bosfer}
\eeq
where $\Lambda$  is the set of all partitions. The identity \eqref{eq:bosfer} is equivalent to Jacobi triple product formula, it can be derived using the fermionic realization of 
$\widehat{u(1)}$
current algebra.  It is less-known that the identity \eqref{eq:bosfer} has a refined version,related to the $W_{1+\infty}$-representation theory \cite{Frenkel:1994em, Awata:1994xm}. Even less-known is the noncommutative generalization which we prove below. 

We proceed with a lemma:

\subsection{Boson-fermion correspondence}

Let $S$ denote the set of all finite subsets of the set ${\BZ}_{\geq 0} + \frac 12$ of positive half-integers. 

{\bf Lemma.}

The product $S \times S$ is in one-to-one correspondence with ${\BZ} \times {\Lambda}$, where we recall that $\Lambda$ is the set of all partitions. The bijection 
\beq
S \times S \leftrightarrow  {\BZ} \times {\Lambda}
\label{lemma:bij}
\eeq 
is established as follows. 

\begin{enumerate} 

\item{} ${\BZ} \times {\Lambda} \to S\times S$: given $M \in \BZ$, $\lambda \in \Lambda$, 
define the sets $({\Sigma}_{+}, {\Sigma}_{-})$ via:
\beq
\begin{aligned}
& {\Sigma}_{+} = \left\{ M + {\lambda}_{i}- i + \frac 12 \, | \, i \in {\BZ}_{> 0} \right\} \cap {\BZ}_{\geq 0} + \frac 12 \, \\
& {\Sigma}_{-} = \left\{ - M + {\lambda}_{j}^{t} - j + \frac 12 \, | \, j \in {\BZ}_{>0} \right\} \cap {\BZ}_{\geq 0} + \frac 12 \, , \\
\end{aligned}
\label{eq:twosets}
\eeq
equivalently
$-{\Sigma}_{-} = {\BZ}_{<0} + \frac 12 \backslash \left\{ M + {\lambda}_{i}- i + \frac 12 \, | \, i \in {\BZ}_{> 0} \right\}$. 
We define 
\beq
d_{\pm} (M, {\lambda}) : = | {\Sigma}_{\pm} |
\label{eq:dpm}
\eeq
\item{} $S \times S \to {\BZ} \times {\Lambda}$: 
given $({\Sigma}_{+}, {\Sigma}_{-})$, define  
\beq
{\psi}_{{\Sigma}_{+}, {\Sigma}_{-}}(u) = \frac{1}{u^{\frac 12}- u^{-\frac 12}}  + \sum_{r_{+} \in {\Sigma}_{+}} u^{r_{+}} - \sum_{r_{-} \in {\Sigma}_{-}} u^{-r_{-}}\, , 
\label{eq:spmfromml}
\eeq
define $M$ by
\beq
M = |{\Sigma}_{+}| - |{\Sigma}_{-}|\, , 
\eeq
then extract $\lambda_{i}$'s from the expansion of \eqref{eq:spmfromml} at $u = \infty$:
\beq
 {\psi}_{{\Sigma}_{+}, {\Sigma}_{-}}(u)  =  \sum_{i=1}^{\infty} u^{M+{\lambda}_{i} - i  +\frac 12} \, , 
\label{eq:uinf}
\eeq
equivalently, extract $\lambda^{t}$ from the expansion at $u = 0$:
\beq
{\psi}_{{\Sigma}_{+}, {\Sigma}_{-}}(u)  =  \sum_{j=1}^{\infty} u^{M+j - {\lambda}_{j}^{t}  -\frac 12} \,  
\label{eq:u0}
\eeq
equivalently, order the set ${\Sigma}_{+} \cup \{ r \, | \, r-\frac 12 \in {\BZ}, \, r< 0 \, , \ - r \notin {\Sigma}_{-} \} = \{ M + {\lambda}_{i} - i + \frac 12 \, | \, i \in {\BN} \}$.
It follows:
\beq
\left( u^{\frac 12}- u^{-\frac 12} \right) {\psi}_{{\Sigma}_{+}, {\Sigma}_{-}}(u)  = u^{M} \left( 1 - (1-u)(1-u^{-1} ) \sum_{(i,j) \in {\lambda}} u^{j-i} \right)  \, . \label{eq:sfrpsi}
\eeq
For future use define the integers $n_{i}$, ${\tilde n}_{j}$ by
\beq
\begin{aligned}
& n_{i} = M + {\lambda}_{i} - i \geq 0 \, , \ i = 1, \ldots , d_{+} \\
& {\tilde n}_{j} = -M + {\lambda}_{j}^{t} - j +1 > 0 \, , \ j = 1, \ldots , d_{-}   \ .
\end{aligned}
\label{eq:lamntn}
\eeq 
obeying $n_{1} > n_{2} > \ldots  > n_{d_{+}} \geq 0$, ${\tilde n}_{1} > \ldots > {\tilde n}_{d_{-}} > 0$.  In terms of $n$'s and ${\tilde n}$'s the half-integers $r_{\pm} \in {\Sigma}_{\pm}$ are given by a simple shift by $\pm \frac 12$, cf. \eqref{eq:twosets}
\beq
r_{+} = n + \frac 12\, , \ r_{-} = {\tilde n} - \frac 12
\eeq
\end{enumerate}
{\bf Remarks}. 


It follows from \eqref{eq:lamntn} that $d_{\pm} = d_{\pm}(M, {\lambda})$ obey
\beq
\begin{aligned}
& {\lambda}_{d_{+}} \geq d_{-} \geq {\lambda}_{d_{+}+1} \\
& {\lambda}_{d_{-}}^{t} \geq d_{+} \geq {\lambda}_{d_{-}+1}^{t} \\
\end{aligned}
\label{eq:bounds}
\eeq
The pair $(d_{+}, d_{-})$ belongs to a set ${\sf S}_{\lambda}$, which we call a snake of $\lambda$. The snake is a union of four sets:

\beq
{\sf S}_{\lambda} = {\sf S}_{\lambda}^{(1)} \, \amalg\, {\sf S}_{\lambda}^{(2)} \, \amalg \, {\sf S}_{\lambda}^{(3)} \, \amalg \, {\sf S}_{\lambda}^{(4)}
\label{eq:snake}
\eeq

\begin{enumerate}

\item{}

${\sf S}_{\lambda}^{(1)}$: 
Either $d_{-} = 0$ and $d_{+} > {\lambda}_{1}^{t}$, or both $d_{+}, d_{-} > 0$ and ${\lambda}_{d_{-}}^{t} > d_{+} > {\lambda}_{d_{-}+1}^{t}$. Moreover
\beq
{\forall} \, i \, , \ d_{+} \geq i > {\lambda}_{d_{-}+1}^{t}, \qquad {\lambda}_{i} = d_{-}
\label{eq:vertsnake}
\eeq

\item{}

${\sf S}_{\lambda}^{(2)}$: 
Either $d_{-} = 0$ and $d_{+} = {\lambda}_{1}^{t}$, or $d_{+} = 0$ and $d_{-} = {\lambda}_{1}$, or both $d_{+}, d_{-} > 0$ and ${\lambda}_{d_{-}}^{t} > d_{+} = {\lambda}_{d_{-}+1}^{t}$, ${\lambda}_{d_{+}} > d_{-} = {\lambda}_{d_{+}+1}$

\item{}

${\sf S}_{\lambda}^{(3)}$: 
Either $d_{+} = 0$ and $d_{-} > {\lambda}_{1}$, or both $d_{+}, d_{-} > 0$ and 
${\lambda}_{d_{+}} > d_{-} > {\lambda}_{d_{+}+1}$. Moreover
\beq
{\forall} \, j \, , \ d_{-} \geq j > {\lambda}_{d_{+}+1}, \qquad {\lambda}_{j}^{t} = d_{+}
\label{eq:horizsnake}
\eeq

\item{}

${\sf S}_{\lambda}^{(4)}$: 
Both $d_{+}, d_{-} > 0$ and ${\lambda}_{d_{+}} = d_{-} > {\lambda}_{d_{+}+1}$, ${\lambda}_{d_{-}}^{t} = d_{+} > {\lambda}_{d_{-}+1}^{t}$.

\end{enumerate}
${\blacksquare}$

We now generalize the identity \eqref{eq:bosfert} to the non-commutative setting, with factors
taking values in a non-commutative algebra $\CalA$ containing a commutative subalgebra $\CalC$. 
We pick a subset of $\CalC$, organized into an $\infty \times \infty$ matrix ${\bf Y} = \Vert Y_{a,b} \Vert_{a,b \in {\BZ}}$. We assume there exists an element ${\bf S} \in {\CalA}$, obeying
\beq
{\bf S} Y_{a,b} {\bf S}^{-1} = Y_{a+1, b+1}
\label{eq:sysy}
\eeq
Define, for $\lambda \in \Lambda$:
\beq
{\CalX}_{\lambda} ({\bf Y}) = Y_{0,0} \prod_{(a,b) \in {\lambda}} \frac{Y_{a-1,b} }{Y_{a-1,b-1}} \frac{Y_{a,b-1}}{Y_{a,b}}
\label{eq:xlamy}
\eeq
and
\beq
{\CalX} ({\bf Y} ) = \sum_{\lambda} {\CalX}_{\lambda} ({\bf Y})
\label{eq:xy}
\eeq
Define, for $M \in {\BZ}$, 
\beq
^{[-M]}{\bf Y} = \Vert Y_{a-M,b} \Vert_{a, b \in {\BZ}}\, , \ {\bf Y}^{[-M]} = \Vert Y_{a,b-M} \Vert_{a, b \in {\BZ}}\
\eeq
Then, we have 
\beq
\begin{aligned}
& The\ noncommutative\ Jacobi\  identities: \\
& \qquad {\CalZ} ({\bf Y}) = \sum_{M \in {\BZ}} {\CalX} (^{[M]}{\bf Y})\ {\bf S}^{M} \in {\CalA} \\
& \qquad {\CalZ}^{\sf T} ({\bf Y}) = \sum_{M \in {\BZ}} {\CalX} ({\bf Y}^{[M]})\ {\bf S}^{M} \in {\CalA}\\
\label{eq:mainJ1}
\end{aligned}
\eeq
where (assuming convergence in some topology)
\begin{multline}
{\CalZ} ({\bf Y}) = \overleftarrow{\prod_{n=1}^\infty} \Big(1 + \frac{Y_{n,0}}{Y_{n,1}}  \, {\bf S} \Big) 
\cdot Y_{0,0} \cdot \overrightarrow{\prod_{n=0}^\infty} \Big(1 + \frac{Y_{-1,n}}{Y_{0,n}} {\bf S}^{-1} \Big) \, , \\
{\CalZ}^{\sf T} ({\bf Y}) = \overleftarrow{\prod_{n=0}^\infty} \Big(1 + {\bf S} \frac{Y_{-1,n}}{Y_{0,n}}  \,  \Big) 
\cdot Y_{0,0} \cdot \overrightarrow{\prod_{n=1}^\infty} \Big(1 + {\bf S}^{-1} \frac{Y_{n,0}}{Y_{n,1}}  \Big) 
\label{eq:mainJ2}
\end{multline}

{\bf Proof.} ${\CalX}_{\lambda}({\bf Y}^{[-M]})$ is a product over $\Sigma_{+} \times {\Sigma}_{-}$, cf. \eqref{lemma:bij}:
\begin{multline} 
{\CalX}_{\lambda}(^{[-M]}{\bf Y}) \, = \, 
Y_{-M,0} \prod_{(i,j) \in {\lambda}} \frac{Y_{-M+i-1,j} \, Y_{-M+i,j-1}}{Y_{-M+i-1,j-1} \, Y_{-M+i,j}} = \\
= Y_{d_{-}, d_{-}} \prod_{j=1}^{d_{-}} \left(    \frac{Y_{-M + {\lambda}_{j}^{t}, j-1}}{Y_{-M+{\lambda}_{j}^{t}, j}} \right) \cdot 
\prod_{i=1}^{d_{+}} \left( \frac{Y_{-M+i-1, {\lambda}_{i}}}{Y_{-M+i, {\lambda}_{i}}} \right) = \\
{\bf S}^{d_{-}} Y_{0,0} {\bf S}^{-d_{-}} \prod_{j=1}^{d_{-}} {\bf S}^{j-1} \left(    \frac{Y_{{\tilde n}_{j},0}}{Y_{{\tilde n}_{j}, 1}} \right) {\bf S}^{1-j} \cdot 
\prod_{i=1}^{d_{+}} {\bf S}^{-M+i} \left( \frac{Y_{-1, n_{i}}}{Y_{0, n_{i}}} \right)
{\bf S}^{M-i} = \\
\overrightarrow{\prod_{j=1}^{d_{-}}} \left(  \frac{Y_{{\tilde n}_{j},0}}{Y_{{\tilde n}_{j}, 1}}  {\bf S} \right) \cdot Y_{0,0} \cdot \overleftarrow{\prod_{i=1}^{d_{+}}} \left( \frac{Y_{-1, n_{i}}}{Y_{0, n_{i}}} {\bf S}^{-1} \right) \cdot {\bf S}^{M}
\label{eq:split}
\end{multline}
where $d_{\pm} = |{\Sigma}_{\pm}|$, $M = d_{+}- d_{-}$, as before. 

A comment on the passage from the first to the second line in \eqref{eq:split}. If both $d_{+}=d_{-}=0$ then $\lambda = {\emptyset}$ and \eqref{eq:split} is obvious. If one of the $d$'s is positive, say $d_{-}$, then partition the set of $(i,j) \in {\lambda}$ into two subsets: ${\lambda}_{-} \subset {\lambda}$, consisting of pairs $(i,j)$, where $1 \leq j \leq d_{-}, 1 \leq i \leq {\lambda}_{j}^{t}$, and ${\lambda}_{+} \subset {\lambda}$, consisting of pairs $(i,j)$, where ${\lambda}_{i} \geq j > d_{-}$, and  $1 \leq i \leq {\lambda}_{d_{-}+1}^{t} \leq d_{+}$, cf. \eqref{eq:bounds}.   The product \eqref{eq:split} then factorizes as
\beq
Y_{-M,0} \prod_{(i,j) \in {\lambda}_{-}} \frac{Y_{-M+i-1,j} \, Y_{-M+i,j-1}}{Y_{-M+i-1,j-1} \, Y_{-M+i,j}} = Y_{-M,d_{-}} \prod_{j=1}^{d_{-}} \frac{Y_{-M+{\lambda}_{j}^{t},j-1}}{Y_{-M+{\lambda}_{j}^{t},j}} 
\label{eq:factor1}\eeq
times
\beq
\prod_{i=1}^{{\lambda}_{d_{-}+1}^{t}} \prod_{j=d_{-}+1}^{{\lambda}_{i}}  \frac{Y_{-M+i-1,j} \, Y_{-M+i,j-1}}{Y_{-M+i-1,j-1} \, Y_{-M+i,j}}
= \frac{Y_{-M+{\lambda}_{d_{-}+1}^{t},d_{-}}}{Y_{-M,d_{-}}}\, \prod_{i=1}^{{\lambda}_{d_{-}+1}^{t}} \frac{Y_{-M+i-1,{\lambda}_{i}}}{Y_{-M+i,{\lambda}_{i}}} \, .
\label{eq:factor2}
\eeq
Recalling \eqref{eq:vertsnake} we write:
\beq
 Y_{-M+{\lambda}_{d_{-}+1}^{t},d_{-}} = Y_{d_{-}, d_{-}} \prod_{i={\lambda}_{d_{-}+1}^{t}+1}^{d_{+}} \frac{Y_{-M+i-1,{\lambda}_{i}}}{Y_{-M+i,{\lambda}_{i}}}
 \label{eq:factor3}
 \eeq
 Substituting \eqref{eq:factor3} into \eqref{eq:factor2} and multiplying by \eqref{eq:factor1}
 we arrive at the second line in \eqref{eq:split}. 
 
 The case of $d_{+} > 0$ is treated analogously, with the help of \eqref{eq:horizsnake}. 
 The second identity is proven by replacing ${\bf Y}$ with ${\bf Y}^{\sf T}$.

$\blacksquare$

\subsection{Non-commutative Hirota identity}

Suppose now we are given two infinite matrices 
\beq
{\bf Y} = \Vert Y_{a,b} \Vert \, , \ {\tilde {\bf Y}} = \Vert {\tilde Y}_{a,b} \Vert \,  , \ a,b \in \BZ .
\eeq
with $Y_{a,b}, {\tilde Y}_{a,b} \in {\CalC}$, such that
\beq
{\bf S} Y_{a,b} {\bf S}^{-1} = Y_{a+1, b+1}\, , \ {\bf S} {\tilde Y}_{a,b} {\bf S}^{-1} = {\tilde Y}_{a-1,b}
\label{eq:sysyt}
\eeq
Which are connected via:
\beq
\frac{{\tilde Y}_{1,n}}{{\tilde Y}_{0,n}} = \frac{Y_{1,n}}{Y_{0,n}}\, , \ n \in {\BZ}
\label{eq:connectyyt}
\eeq
{\bf Theorem} the following \emph{bilinear identity}
\beq
\sum_{{\lambda}, {\mu} \in {\Lambda}, \, M \in {\BZ}} (-1)^{M} {\CalX}_{\lambda} \left[ ^{[-M]}{\bf Y} \right]
{\CalX}_{\mu} \left[ ^{[M+1]}{\bf{\tilde Y}} \right] = 0 
\label{eq:nchirota}
\eeq
holds (cf. \eqref{eq:xlamy}). The proof uses the remarkable involution ${\rho}$ acting on ${\Lambda} \times {\BZ} \times {\Lambda}$, found in \cite{NG}
\beq
{\rho}({\lambda}, M, {\mu}) = \begin{cases} {\mu}_{1} - {\lambda}_{1} > M \, , & \ ({\tilde\lambda}, M+1, {\tilde\mu})
\, , \\
& \ {\rm where}\ {\tilde\lambda}_{1} = {\mu}_{1} - M - 1 \, , \ \\
& \ {\tilde\lambda}_{a+1} = {\lambda}_{a}\, , {\tilde\mu}_{a} = {\mu}_{a+1}\, , \ a \geq 1 \,  , \\
{\mu}_{1} - {\lambda}_{1} \leq M\, , & \ ({\hat\lambda}, M-1 , {\hat\mu}) \, , \\
& \ {\rm where} \ {\hat\mu}_{1} = {\lambda}_{1} + M  \\
& \ {\hat\lambda}_{a} = {\lambda}_{a+1} \, , \ {\hat\mu}_{a+1} = {\mu}_{a}\, , a \geq 1 \, , \\
\end{cases}
\eeq
It is easy to compute, for ${\mu}_{1} - {\lambda}_{1} > M$
\begin{multline}
\frac{{\CalX}_{\tilde\lambda}\left[ ^{[-M-1]}{\bf Y} \right]{\CalX}_{\tilde\mu}\left[ ^{[M+2]}{\bf {\tilde Y}} \right]}{{\CalX}_{\lambda}\left[ ^{[-M]}{\bf Y} \right]{\CalX}_{\mu}\left[ ^{[M+1]}{\bf {\tilde Y}} \right]} \, = \\
 \frac{Y_{-M-1,{\mu}_{1}-M-1}}{Y_{-M,{\mu}_{1}-M-1}} \frac{{\tilde Y}_{M+2,{\mu}_{1}}}{{\tilde Y}_{M+1, {\mu}_{1}}} = 
{\bf S}^{-M-1} \left( \frac{Y_{0,{\mu}_{1}}}{Y_{1,{\mu}_{1}}} \frac{{\tilde Y}_{1,{\mu}_{1}}}{{\tilde Y}_{0,{\mu}_{1}}} \right) {\bf S}^{M+1} = 1
\end{multline}
using \eqref{eq:sysyt} and \eqref{eq:connectyyt}. Thus, we match the terms in \eqref{eq:nchirota} corresponding to 
the triples $({\lambda}, M, {\mu})$ and $({\tilde\lambda}, M+1, {\tilde\mu})$, thereby ensuring the cancellation. $\blacksquare$

\section{First applications}

We consider several examples:

\subsection{$W_{1+{\infty}}$-Jacobi identity}

Let $\bf Y$ be a Toeplitz matrix, i.e. its entries commute with $\bf S$, 
\beq
Y_{a,b} = e^{{\sf b}(b-a)}\, , \label{eq:bij}
\eeq 
where ${\sf b}({\xi}) = \sum_{k=1}^{\infty}  \frac{b_{k}}{k!} {\xi}^k$ with $b_{k}$, $k>2$
 formal variables. Define $t({\xi})$ by
\beq
t({\xi}) = {\sf b}({\xi} + \frac 12 ) - {\sf b}({\xi} - \frac 12 )\ . 
\label{eq:tfb}
\eeq
Then \eqref{eq:mainJ1} specifies 
to the following generalization of Jacobi triple product identity:
\begin{multline}
\prod_{r > 0} \left( 1 +  e^{t(r)} \right) \left( 1 + e^{-t(-r)} \right) 
 = \\
 \sum_{M\in {\BZ}} e^{{\sf b}(M)} \sum_{\lambda} {\qe}^{|{\lambda}|} \, \prod_{i=1}^{\infty} e^{t^{\rm formal} \left( M + \frac 12 + {\lambda}_{i} - i \right) - t^{\rm formal}  \left( M + \frac 12 - i \right)}
\label{eq:bosfert}
\end{multline}
where
${\qe} = e^{b_{2}}$ should obey $|{\qe}| < 1$ for convergence, and 
\beq
t^{\rm formal} = \sum_{k, l \geq 0\, , k+2l > 1} \frac{b_{k+2l+1}}{2^{2l} (2l+1)!k!} {\xi}^{k}
\eeq
In the particular case $b_{k} = 0$, $k >3$, the identity \eqref{eq:bosfert} can be found  in \cite{Dijkgraaf:1996iy}. In a slightly different form the identity \eqref{eq:bosfert} is discussed in \cite{BO}.

\subsection{$\Theta$-transform of $q$-characters}

In these example the matrix $\bf Y$ is defined in terms of the $Y$-observables of ${\hat A}_{r}$-type quiver gauge theory in four dimensions. The reader might want to consult \cite{BPSCFT1} for explanations of some of the notations. 

\subsubsection{Lightning review of instanton counting}

To start, 
the setup of $\Omega$-deformed $\CalN = 2$ gauge theory requires  
a $4$-tuple $\boldsymbol{\ve}$ of complex numbers ${\ve}_{a}$, $a = 1, \ldots, 4$, s.t.
\beq
{\ve}_{1} + {\ve}_{2} +{\ve}_{3} + {\ve}_{4} = 0
\label{eq:sl4}\eeq
We use the notations:
\beq
{\ve} = {\ve}_{1}+{\ve}_{2}\, , \ {\tilde\ve}_{3} = \frac{\ve_3}{r+1}\, , \ {\tilde\ve}_{4} = - {\ve} - {\tilde\ve}_3\, .  \\
\label{eq:epsparam}
\eeq
Fix an integer $r \geq 0$. We are going to study the four dimensional gauge theory with the gauge group $U(N)^{r+1}$, 
characterized by the complexified couplings ${\qe}_{i} = {\exp}\, 2\pi \ii \tau_i$, 
\beq
{\tau}_{i} = \frac{{\vt}_{i}}{2\pi} + \frac{4\pi\ii}{g_{i}^{2}}\, , \ i = 0, \ldots, r
\label{eq:coupl}
\eeq
with the combinations
\beq
{\qe} = \prod_{i=0}^{r} {\qe}_{i}\, , \ 
{\tilde\qe} = {\qe}^{\frac 1{r+1}}
\eeq
playing a special role. 
To each gauge group factor $U(N)_{i}$ we associate a collection of complex numbers ${\bf a}_{i} = (a_{i, \alpha})_{\alpha = 1}^{N} \in {\BC}^{N}$, called the \emph{vev of the complex scalar in the $i$'th adjoint representation}, or, 
\emph{equivariant parameters for the framing symmetry} for short. We denote ${\bf a} = ( {\bf a}_{i} )_{i=0}^{r} = ( a_{i, \alpha} )_{0 \leq i \leq r\, , \, 1 \leq {\alpha} \leq N}\in {\BC}^{N(r+1)}$.

We define the
$Y$-observables $y_{i}(x)$ as functions on the infinite set of $N(r+1)$-tuples
${\bla} = ({\lambda}^{(i,{\alpha})})$, $i = 0, \ldots, r$, ${\alpha} = 1, \ldots, N$ of Young diagrams (partitions), defined by
\beq
y_{j}(x) \biggr\vert_{\bla} = \prod_{\alpha = 1}^{N} \left( x - a_{j, \alpha} \right) \times
\frac{K_{j}(x-{\ve}_{1})  K_{j}(x-{\ve}_{2}) }{K_{j}(x)  K_{j}(x-{\ve}_{1}-{\ve}_{2}) } \biggr\vert_{{\bla}}
\eeq
where
\beq
K_{j}(x) \biggr\vert_{\bla} = \prod_{\alpha=1}^{N} \prod_{(a,b) \in {\lambda}^{(j,{\alpha})}}
(x - a_{j, \alpha} - {\ve}_{1}(a-1) - {\ve}_{2}(b-1))
\eeq
The value of the $Y$-observable on the \emph{random variable} ${\bla}$ is a rational function of $x$. The expectation value of observables is defined as a sum over all $\bla$
of the evaluation of observable at $\bla$ times the certain measure factor 
\beq
m_{\bla}( {\bf a},{\boldsymbol{\ve}}; {\boldsymbol{\tau}}) = \, \prod_{i=0}^{r} \prod_{\alpha=1}^{N} {\qe}_{i}^{-\frac{a_{i, \alpha}^{2}}{2{\ve}_{1}{\ve}_{2}} + | {\lambda}^{(i, {\alpha})} |} \, \times \, {\rm rational\ function\ of }\ {\bf a} \  {\rm and}\ {\boldsymbol{\ve}}
\eeq
\beq
\langle {\CalO} \rangle = \frac{1}{{\CalZ} ( {\bf a},{\boldsymbol{\ve}}; {\boldsymbol{\tau}} )} \sum_{\bla} \, {\CalO} \biggr\vert_{\bla}\, m_{\bla}( {\bf a},{\boldsymbol{\ve}}; {\boldsymbol{\tau}}) 
\label{eq:expval}
\eeq
divided by the partition function
\beq
{\CalZ} ( {\bf a},{\boldsymbol{\ve}}; {\boldsymbol{\tau}} ) = \sum_{\bla} m_{\bla}( {\bf a},{\boldsymbol{\ve}}; {\boldsymbol{\tau}})
\eeq
The algebra ${\CalA}$ can be realized as difference operators (possibly, of infinite order)
acting on ${\BC}^{r+1}$-valued analytic functions of the variable $x$ (we don't specify precise analytic conditions on the functions), which we view as the space of quasiperiodic functions ${\psi}_{i}(x)$ on ${\BZ} \times {\BC}$, obeying
\beq
{\psi}_{i+r+1}(x) = {\psi}_{i}(x+{\ve}_{3})
\label{eq:qper1}
\eeq
The commutative subalgebra $\CalC$ is the algebra of diagonal $r+1 \times r+1$ matrices-valued functions
of $x$. 
Let us extend the defintion of the couplings to $i \in {\BZ}$ by
\beq
{\tau}_{i+r+1}\, = \, {\tau}_{i} \, , 
\eeq
and define $\xi_{i} = \xi_{i+r+1}$ as any solution to the difference Laplace equation:
\beq
{\xi}_{i-1} - 2 {\xi}_{i} + {\xi}_{i+1} \, = \,  {\rm log}\left( {\qe}_{i} {\tilde\qe}^{-1} \right)\ .
\label{eq:etas}
\eeq

\subsubsection{$\bf Y$-matrix from $Y$-observables }
For fixed $i$, define the matrix $\bf Y$ (cf. \eqref{eq:expval}) by:
\beq
Y_{a,b} =  {\tilde\qe}^{- \frac{\left(x+{\ve}(1-b)+(i+a-b){\tilde\ve}_{3}\right)^2}{2{\tilde\ve}_{3}{\tilde\ve}_{4}}} e^{{\xi}_{i+a-b}} \, \Biggl\langle y_{i+a-b}\left( x+{\ve}(1-b) \right) \Biggr\rangle \, 
\label{eq:Ym34}
\eeq
The operator ${\bf S}$ is simply the shift
\beq
{\bf S} = e^{-{\ve}{\partial}_{x}}
\label{eq:S34}
\eeq
Applying \eqref{eq:mainJ1} to \eqref{eq:Ym34} we obtain, in the limit ${\ve}_{1} \to 0$ with ${\ve}_{2} = {\ve}$ kept fixed (or, equivalently, ${\ve}_{2} \to 0$ with ${\ve}_{1} = {\ve}$ fixed) the relation between the expectation values of the $q$-characters and the $y$-observables of the ${\hat A}_{r}$-type quiver gauge theory 

\subsubsection{Factorized $\Theta$-transform of $q$-characters}

\begin{multline}
{\tilde{\CalD}}_{i} : = \sum_{M \in {\BZ}}  
{\tilde\qe}^{- \frac{\left(x+{\ve}+(i-M){\tilde\ve}_{3}\right)^2}{2{\tilde\ve}_{3}{\tilde\ve}_{4}}}  e^{{\xi}_{i-M}}\, {\chi}_{i-M} (x)  e^{{\ve}M {\partial}_{x}}  \, = \, 
{\tilde{\CalD}}_{i}^{+}  \cdot {\tilde\qe}^{- \frac{\left(x+{\ve}+i{\tilde\ve}_{3}\right)^2}{2{\tilde\ve}_{3}{\tilde\ve}_{4}}}  e^{{\xi}_{i}}\, y_{i}(x+{\ve}) \cdot {\tilde{\CalD}}_{i}^{-}\\
{\tilde{\CalD}}_{i}^{+} : = \overleftarrow{\prod_{n=1}^\infty} \Big(1 +   {\tilde\qe}^{\frac{x - \frac{\tilde\ve_4}{2}}{{\tilde\ve}_{3}}} \,  {\tilde\qe}^{i+n-1} \, e^{{\xi}_{i+n}-{\xi}_{i+n-1}} \frac{y_{i+n}(x+{\ve})}{y_{i+n-1}(x)}   \cdot e^{-{\ve}{\partial}_{x}} \Big) \, , 
\\
{\tilde{\CalD}}_{i}^{-} : =
\overrightarrow{\prod_{n=0}^\infty} \Big(1 + e^{{\ve}{\partial}_{x}} \cdot  {\tilde\qe}^{\frac{\left( x+ \left( i-  \frac 12 \right) {\tilde\ve}_{3}\right)}{{\tilde\ve}_{4}}}  {\tilde\qe}^{n} e^{{\xi}_{i-n-1}-{\xi}_{i-n}} \, \frac{y_{i-n-1}(x-n{\ve})}{y_{i-n}(x-n{\ve})}  \Big) \, . 
\label{eq:MainJ34}
\end{multline}
First of all, even though we have identities like \eqref{eq:MainJ34} for any $i \in \BZ$, the operators ${\CalD}_{i}$
are all related:
\beq
{\tilde\CalD}_{i} = {\tilde\CalD} \, {\bf S}^{-i}
\label{eq:cdi}
\eeq
where ${\tilde\CalD} := {\tilde\CalD}_{0}$. 
In addition, the twisted periodicity implies:
\beq
e^{{\ve}_{3}{\partial}_{x}} \, {\tilde\CalD}  \, = \, {\tilde\CalD}\,  e^{-{\ve}_{4}{\partial}_{x}}
\eeq
In other words, $\tilde\CalD_0$ intertwines the $x$-shifts by ${\ve}_{3}$ and by $- {\ve}_4 = {\ve}_{3}+(r+1){\ve}$.

\subsubsection{Convergence of sums and products}

Let us now analyze the convergence of both the left and the right hand sides of \eqref{eq:MainJ34}. 

The factors in ${\tilde\CalD}_{i}^{\pm}$ of \eqref{eq:MainJ34} behave as  $1 + {\tilde\qe}^{n} \times $ some functions of $(x, i)$, making the convergence of the product over $n$ not obvious.  
The left hand side contains the factors
$\propto {\tilde\qe}^{- \frac{\tilde\ve_3}{2\tilde\ve_4} M^2}$, making the convergence of the sum over $M\in \BZ$ quite questionable. However, there is a simple way of transforming \eqref{eq:MainJ34} into the manifestly convergent sum and product by multiplying ${\tilde\CalD}_{i}$ by appropriate functions both on the left and on the right:
\beq
{\CalD}_{i} :=  e^{-{\xi}_{i}} \, {\tilde\qe}^{-\frac{\left( x + {\ve} + i{\tilde\ve}_{3} \right)^{2}}{2 {\ve} {\tilde\ve}_3}} {\tilde{\CalD}}_{i} {\tilde\qe}^{-\frac{\left( x + {\ve} + i{\tilde\ve}_{3}\right)^{2}}{2 {\ve} {\tilde\ve}_4}} = {\CalD}_{i}^{+} y_{i}(x+{\ve}) {\CalD}_{i}^{-}
\eeq
Using the identity
\beq \label{Fay}
- \frac{1}{{\tilde\ve}_{3}{\tilde\ve}_{4}} = \frac{1}{{\tilde\ve} {\tilde\ve}_{4}} + \frac{1}{{\tilde\ve} {\tilde\ve}_{3}}
\eeq
we arrive at:
\begin{multline}
{\CalD}_{i}  = \sum_{M \in {\BZ}}   \, \tilde{\mathfrak{q}}^{\frac{M^2}{2}}e^{\xi_{i-M} - \xi_i} \, {\chi}_{i-M} (x)  e^{{\ve}M {\partial}_{x}}  = {\CalD}_{i}^{+} y_{i}(x+{\ve}) {\CalD}_{i}^{-}\, , \\
{\CalD}_{i}^{+} = 
\overleftarrow{\prod_{n=1}^\infty} \Big(1 +   \tilde{\mathfrak{q}}^{n-\frac{1}{2}}e^{\xi_{i+n} - \xi_{i+n-1}} \frac{y_{n}(x+{\ve})}{y_{n-1}(x)}   \cdot e^{-{\ve}{\partial}_{x}} \Big) 
\, , \\
{\CalD}_{i}^{-} = 
\overrightarrow{\prod_{n=0}^\infty} \Big(1 + e^{{\ve}{\partial}_{x}} \cdot  \tilde{\mathfrak{q}}^{n+\frac{1}{2}}e^{\xi_{i-n-1} - \xi_{i-n}} \, \frac{y_{i-n-1}(x-n{\ve})}{y_{i-n}(x-n{\ve})}  \Big)  
\label{eq:MainJ34i}
\end{multline}
with convergent left and right hand sides.

\section{Gauge theory interpretation}

The significance of \eqref{eq:MainJ34i} is the remarkable property of the $q$-characters ${\chi}_{i}(x)$
(which are the $\ve_2 \to 0$ limits of the expectation values of the $qq$-characters \cite{BPSCFT1}) -- they are polynomials in $x$ of degree $N$. Thus, 
on the small phase space the left hand side of \eqref{eq:MainJ34} is an infinite order difference operator
with polynomial coefficients. The factorization formula given by the right hand side of \eqref{eq:MainJ34} opens a
way to connecting the difference operators, or, via Fourier transform, $SL_N$-opers on $r+1$-punctured elliptic curve ${\BC}^{\times}/{\qe}^{\BZ}$ and the expectation values of the $Y_{i}$-observables. 

In this paper we only briefly discussed the $Y_{i}$-observables, presenting their combinatorial definition. Their true nature is revealed by cohomological definition as homological Chern polynomials of the fibers of universal sheaves on the moduli spaces of framed quiver torsion free sheaves on ${\BP}^{2}$ \cite{BPSCFT1}. 

The $\Theta$-transform given by the left hand side of \eqref{eq:MainJ34} can be interpreted in gauge theory language, at least in the classical limit ${\ve}_{1}, {\ve}_{2} \to 0$. In this limit the shift operator
${\bf S}$ becomes a $c$-number $z$ and \eqref{eq:MainJ34} reduces to the identity (whose exploration in the $r=0$ case was initiated in \cite{Nekrasov:2012xe}) 
\begin{multline}
\sum_{M \in {\BZ}} {\chi}_{i-M}(x) z^{M} = \\
\prod_{n=1}^{\infty} \Big(1 +   \tilde{\mathfrak{q}}^{n-\frac{1}{2}}e^{\xi_{i+n} - \xi_{i+n-1}} 
 \frac{y_{i+n}(x)}{y_{i+n-1}(x)}   z^{-1} \Big) 
\cdot  \,
y_{i}(x) \cdot  \,
\prod_{n=0}^{\infty} \Big(1 + z  \tilde{\mathfrak{q}}^{n+\frac{1}{2}}e^{\xi_{i-n-1} - \xi_{i-n}} \, \frac{y_{i-n-1}(x)}{y_{i-n}(x)}  \Big)
\label{eq:red34}
\end{multline}
Let us start with the $r=0$ case. 
Recall \cite{BPSCFT1} that  $\chi_{i}(x) = {\chi}(x-i {\ve}_{3})$ is given by a sum over partitions $\lambda$ (the same sum as in the left hand side of the main identity), which are in one-to-one correspondence with the fixed points of the torus action on the Hilbert scheme of points on ${\BC}^{2}$, which happen to coincide with the moduli space of $U(1)$ instantons on 
noncommutative ${\BR}^{4}$. The ${\BC}^{2}$ in question is usually denoted by ${\BC}^{2}_{34}$, it is not the physical spacetime ${\BC}^{2}_{12}$ of the ${\CalN}=2^{*}$ theory we are studying. Instead, it is an auxiliary space (it is transverse to ${\BC}^{2}_{12}$ in the IIB string construction of gauge origami), supporting an auxiliary $U(1)$ gauge theory (for the fundamental $qq$-character), interacting with the physical $U(N)$ theory on ${\BC}^{2}_{12}$ via bi-fundamental fields, supported in the small neighborhood of the point $0 \times 0 \in {\BC}^{2}_{12} \times {\BC}^{2}_{34}$. The contributions of the fixed points $\lambda$ given by ${\CalX}_{\lambda}[{\bf Y}]$ are precisely the result of integrating out the degrees of freedom, living at the intersection of the physical and auxiliary spaces. The variable $x$, the argument of $\chi$, is the vev of the complex scalar in the vector multiplet of the $U(1)$ gauge symmetry living on ${\BC}^{2}_{34}$. 

Now, let us interpret the left hand side of \eqref{eq:red34}. The sum over $\lambda$ giving rise to $\chi(x)$ is the usual integration over the moduli space of instantons on ${\BC}^{2}_{34}$. What about the sum over $M \in {\BZ}$?  We can shed some light on it by recalling the formulas for the partition functions of ${\CalN}=2$ twisted gauge theories on blowups \cite{Nakajima:2003pg}
or on general toric surfaces \cite{Nekrasov:2006otu}. In these cases the torus fixed points correspond not only to the point-like instantons, located at the fixed points of the torus action on spacetime itself, but also to the line bundles of nontrivial degree, represented by connections whose curvature is essentially localized near the torus invariant curves, ${\BP}^{1}$'s connecting the fixed points in spacetime. For example, for spacetime of the form ${\widehat{{\BC}^{2}}}$, i.e. the ${\BC}^{2}$ with a blown up origin, there is a finite area ${\BP}^{1}$, the exceptional divisor $E$, which can support an arbitrary abelian gauge field flux $M = \frac{1}{2\pi\ii} \int_{E} F \in {\BZ}$. It is the sum over this flux that produces the theta function in the celebrated blowup formula of Fintushel and Stern \cite{FS}, a mathematical precursor of Seiberg-Witten solution of ${\CalN}=2$ theory \cite{SW}. In $\Omega$-deformed theories, i.e. in equivariant version of gauge theory, the additional effect is the shift \cite{Nakajima:2003pg,Nekrasov:2006otu} of the equivariant parameters corresponding to the framing symmetry
by the abelian gauge flux times an appropriate equivariant parameter for the rotational symmetry (an explanation using the effective gauge theory is given in the Eqs. (240-244) in \cite{Nekrasov:2020qcq}). 

In view of these considerations, 
the left hand side of \eqref{eq:red34}
can be now interpreted as the partition function of gauge origami \cite{BPSCFT3}
on the product of ${\BC}^{2}_{12} \times$ a  Taub-Nut space replacing the ${\BC}^{2}_{34}$-factor, with the Chan-Paton spaces $N_{12} = {\BC}^{N}$ and $N_{34} = {\BC}$. The summation over $M \in {\BZ}$ is the feature of gauge theory on Taub-Nut space. The latter has non-trivial $L^2$ cohomology in degree two, even though as a smooth 
manifold it is diffeomorphic to ${\BR}^{4}$. The $U(1)$ gauge theory on Taub-Nut, therefore, involves the summation over the fluxes of the $U(1)$ gauge field, proportional to the normalizable self-dual harmonic two-form (this feature of Taub-Nut space is responsible for the duality between $M$-theory on Taub-Nut space and $D6$-brane in IIA string theory). The gauge theory interpretation of the right hand side of \eqref{eq:red34} is somewhat mysterious. Since the two dimensional perspective on this identity identifies the product with the fermion partition function, it
suggests the presence of fermionic (presumably extended) degrees of freedom in four dimensional gauge theory. The case $r> 0$ can be treated similarly, by replacing the Taub-Nut space but its $r+1$-center version. 

We postpone the interpretation of \eqref{eq:nchirota} till further publications \cite{NGtoap, NGhighertimes}, where we shall apply the identities derived in this paper to classical and quantum integrable systems and the theory of isomonodromy deformations.

\end{document}